# Multilevel Image Encryption

Rakesh S, Ajitkumar A Kaller, Shadakshari B C and Annappa B

Department of Computer Science and Engineering, National Institute of Technology Karnataka, Surathkal
`{rakeshsmysore,ajitkaller,shadsbellekere}@gmail.com,annappa@ieee.org`

## ABSTRACT

*With the fast evolution of digital data exchange and increased usage of multi media images, it is essential to protect the confidential image data from unauthorized access. In natural images the values and position of the neighbouring pixels are strongly correlated. The method proposed in this paper, breaks this correlation increasing entropy of the position and entropy of pixel values using block shuffling and encryption by chaotic sequence respectively. The plain-image is initially row wise shuffled and first level of encryption is performed using addition modulo operation. The image is divided into blocks and then block based shuffling is performed using Arnold Cat transformation, further the blocks are uniformly scrambled across the image. Finally the shuffled image undergoes second level of encryption by bitwise XOR operation, and then the image as a whole is shuffled column wise to produce the ciphered image for transmission. The experimental results show that the proposed algorithm can successfully encrypt or decrypt the image with the secret keys, and the analysis of the algorithm also demonstrates that the encrypted image has good information entropy and low correlation coefficients.*

## 1. Introduction

Encryption is a common technique to uphold multimedia image security in storage and transmission over the network. It has application in various fields including internet communication, medical imaging and military communication. Due to some inherent features of images like high data redundancy and bulk data capacity, the encryption of image differs from that of text, thus algorithms suitable textual data may not be suitable for multimedia data.

Many image-protection techniques use vector quantization (VQ) as the main encryption technique (Chang et al., 2001; Chen and Chang, 2001). A symmetric block encryption algorithm creates a chaotic map, used for permuting and diffusing multimedia image data. There have been many more image encryption algorithms based on chaotic maps [1]-[5]. Also other encryption algorithms based on concepts such as block cipher [6] and selective encryption [9] has been proposed. A few techniques focus on video encryption [10]. Several cryptosystems similar to data encryption, such as steganography [8] and digital signature [7] have also been implemented to increase security of multimedia data storage and transmission.

The approach uses concept of uniform scrambling, row, column and block based image shuffling to reduce correlation. Further, encryption is performed using a chaotic sequence generated by symmetric keys to ensure the security of the proposed algorithm.

## 2. Proposed Method

In order to improve the security of the image encryption algorithm, position of pixels in the original image is shuffled and gray values are encrypted using Logistic Map generated.

## 2.1. Image Encryption Algorithm

Step i.
The plain image I of size NxM is row wise shuffled with a fixed displacement set to the summation of elements in that particular row as shown below-

$$I'(x, y) = I((x + R(x))\%M, y) \quad (1)$$

$$R(x) = \sum_{y=0}^{M} I(x, y) \quad (2)$$

where, I'(x,y) is a pixel value at the co-ordinate (x,y) and R(x) is the sum of all the elements in the $x^{th}$ row of image I.

Step ii.
The essence of image encryption or image scrambling is to reduce the correlation of pixel positions and values until they are irrelevant to each other. Even though the image is shuffled, the pixels will still be having same values. So using entropy and histogram information, attacker can perform statistical attacks making the system vulnerable. So pixel values have to be encrypted to increase entropy. This is done by using the symmetric secret keys A and K to generate chaotic sequence as shown below-

A = K*A*(1–A)
KB = (($10^{14}$)*A)%256
I''(x,y) = (I'(x, y)+KB)%256

where, (x,y) are pixel co-ordinates of the intermediate image I'. Chaotic sequences generates better results when 0<A<1 and 3.5<K<4. In our case, the keys A and K were taken to be 0.3905 and 3.9886 respectively. Above method restricts encrypted values less than 256 using modulus operation, resulting an encrypted image I''.

Step iii.
Now the image is block wise shuffled and scrambled to decrease correlation between adjacent pixels even further. To do this, we divide the image I'' into blocks of size 16x16, $B_1, B_2 \ldots B_n$.

We apply Arnold Cat transformation within each block $B_i$ by the following equation:

$$\begin{bmatrix} x' \\ y' \end{bmatrix} = \begin{bmatrix} 1 & 1 \\ 1 & 2 \end{bmatrix} \begin{bmatrix} x \\ y \end{bmatrix} \quad (3)$$

where (x, y) are original coordinates of I'' and (x',y') are new shuffled coordinates. After repeating this step for each of the n blocks we get partially shuffled image I'''.

Step iv.
Intra-block shuffling would not be sufficient to decrease correlation between the pixels position, they also need to be uniformly scattered across the image. To measure randomness, we consider image position entropy given by the following equation

$$H(I) = \sum_{i=1}^{n} H_i(P) \quad (4)$$

where, n is the total number of image blocks, $H_i(P)$ is the entropy of the $i^{th}$ block and denotes the average information capacity in this block and it is defined as follows

$$H_i(P) = \sum P(x, y) \log \frac{1}{P(x, y)} \quad (5)$$

where, P(x,y) is the probability of pixel with coordinates (x,y) in original image appears at the $i^{th}$ block in the scrambled image. We known that $H_i(P)$ will reach the maximum when all the

P(x,y) are equal and thus H(I) attains the maximum value. So the perfect state of image scrambling is the random pixel in respective block of the scrambled image has the equal probability of coming from the random situation in plain original image. One can also conclude that average information content will get the maximum when the probability that pixels in a block of original image is distribute into different blocks. Thus we perform uniform scrambling where the pixels in the same block of the image I''' is distributed into all the blocks and the every block has one pixel at least, without regarding to the order of the pixels, accordingly all the pixels in the same block of scrambled image come from different blocks of the plain image.

Figure 1 below shows that all the pixels in the first block of the original image are distributed into all the blocks of the scrambled image, irrespective of the order. Thus, ideal block numbers is N for an original image of size NxN. After this uniform scrambling, we get new image IS.

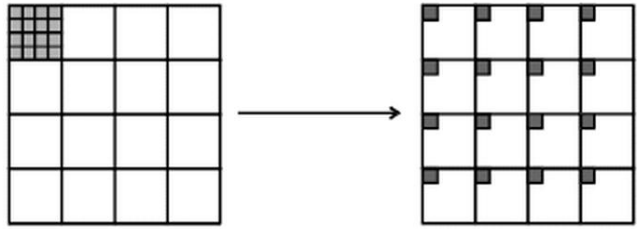

Figure 1. Uniform Image Scrambling

Step v.
Now second level encryption is performed to increase entropy information even further and bring it close to the ideal value of 8. Here pixel values are encrypted using XOR operation with the chaotic sequences generated -

$A = K*A*(1-A)$
$KB = ((10^{14})*A)\%256$
$IS'(x,y) = (IS(x, y) \text{ XOR } KB)\%256$

The resultant will be an encrypted image IS', with entropy close to ideal value. By experimental results, one can see that the histogram of the ciphered image is fairly uniform and is significantly different from that of the plain image, thus not providing any indication to employ statistical attacks on the encrypted image.

Step vi.
Now the cipher image IS' is column wise shuffled with the displacement of a column set to the summation of elements in that particular column as shown below,

$$IE(x,y) = IS'(x, (y + C(y))\%N) \qquad (6)$$

$$C(y) = \sum_{x=0}^{N} I(x,y) \qquad (7)$$

where, IE(x,y) is a pixel value at the co-ordinate (x,y) and C(y) is the sum of all the elements in the y$^{th}$ column. The above steps will result in the final cipher image IE, which could be securely transmitted over the network. Experimental results prove that the proposed approach is resilient against statistical and brute force attacks. Also encryption at different levels using shuffling, scrambling and chaotic mapping make it still harder for an attacker to guess the original image. Image decryption is a simple process of retracing the steps backwards, using the same symmetric key pair A and K to generate chaotic sequence.

Each layer of encryption increases security of the transmitted image a stage higher, but also results in an overhead on computation time. If the application requires a simple encryption mechanism, the first two steps should be sufficient. But in the present age, the proposed approach can be implemented in a multi-core and multiprocessor environment using concepts of pipelining and threading, hence saving on computational time.

## 3. Experimental Results

A good encryption procedure should be robust against all kinds of cryptanalytic, statistical, differential and brute-force attacks. Thus the histogram of the ciphered image must be uniform to avoid statistical attacks, and the key space must be large enough to avoid brute force attacks. Analysis of the proposed approach as shown in figures depicts that it is indeed resilient against possible attacks and also flexible enough to extend the same to binary and RGB images as well.

### 3.1. Histogram Analysis

The plain image, its corresponding cipher image and their histograms are shown in Figure 2. It is clear that the histogram of the cipher image is uniformly distributed and significantly different from the respective histograms of the plain original image. So, the encrypted image does not provide any clue to employ any statistical attack on the proposed procedure, which makes statistical attacks difficult.

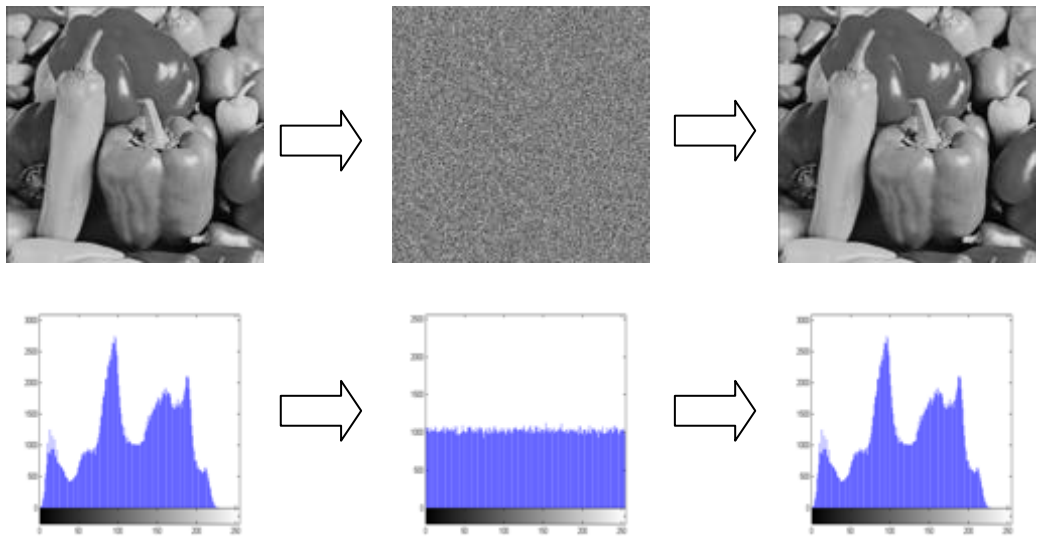

Figure 2. Original, Encrypted and Decrypted Images and their corresponding Histograms

### 3.2. Correlation of two adjacent Pixels

Here, we test the correlation between two vertically and horizontally adjacent pixels in the encrypted image. Correlation coefficient of each pair is calculated by the formula below-

$$cov(p,q) = E(p - E(p))(q - E(q)) \qquad (8)$$

where p and q are pixel values of two adjacent pixels in the image. Fig. (3) (a) shows the distribution of two horizontally adjacent pixels of the plain image, (b) the distribution of two horizontally adjacent pixels of the cipher image, similarly figure (c) shows the distribution of two vertically adjacent pixels of the original image and (d) distribution of two vertically adjacent pixels of the encrypted image.

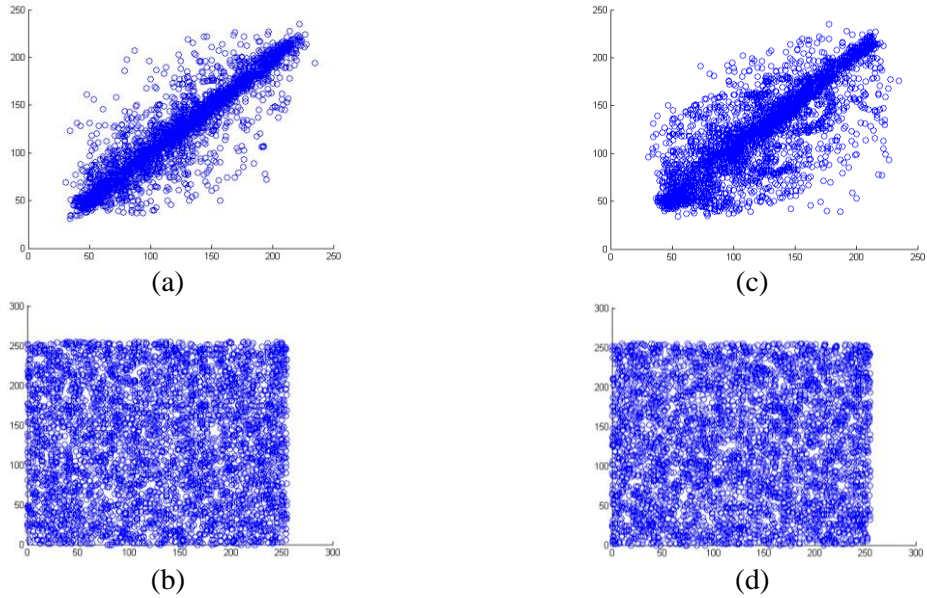

Figure 3. Correlation comparison of two adjacent pixels

### 3.3. Image Entropy

Entropy is a measure of uncertainty association with random variable. As for an image, the encryption decreases the mutual information among pixel values and thus increases the entropy value. A secure system should satisfy a condition on the information entropy that is the cipher image should not provide any information about the original image. The information entropy is calculated using equation

$$Entropy = \sum P(i) \log \frac{1}{P(i)} \qquad (9)$$

where P(i) is the probability of occurrence of a pixel with gray scale value i. If each symbol has an equal probability then entropy of 8 would correspond to complete randomness, which is expected in encrypted image.

Different images have been tested by the proposed image encryption procedure and the resulting entropy, horizontal and vertical correlation coefficients are shown in the Table 1 below.

Table 1. Images with their corresponding Entropy, Horizontal and Vertical Correlation.

| Image | Entropy / Horizontal Correlation / Vertical Correlation | Image | Entropy / Horizontal Correlation / Vertical Correlation |
|---|---|---|---|
| (jet image) | 7.999207 / 0.000962 / -0.001922 | (baboon image) | 7.999253 / 0.000893 / -0.001703 |

| 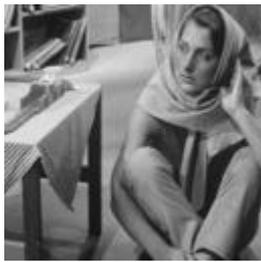 | 7.999227 | 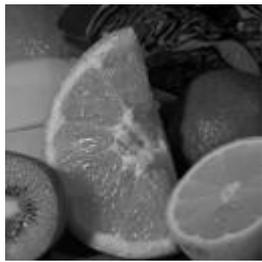 | 7.999342 |
| | -0.001827 | | -0.002541 |
| | 0.001319 | | 0.001462 |
| 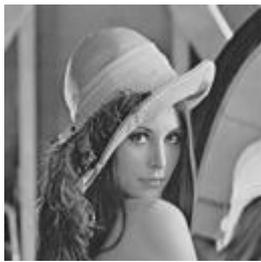 | 7.999308 | 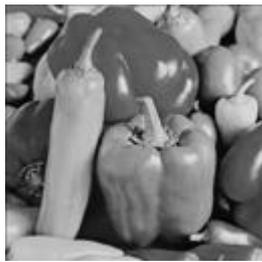 | 7.999951 |
| | 0.000106 | | 0.000308 |
| | 0.001632 | | 0.000312 |

### 3.4. Key Space Analysis

For secure cryptosystem, the key space should be large enough to make sure that brute force attack is infeasible. The proposed algorithm has $2^{256}$ different combinations of the secret keys. A cipher with such a long key space is sufficient for practical use. Furthermore, if we consider the shuffling and scrambling as part of the key, the key space size will be even larger. Hence, the key space of the proposed algorithm is sufficiently large enough to resist the exhaustive of brute-force attacks.

### 3.5. Testing for Special Cases

Table 2 below shows that the proposed algorithm works for some special cases also, as in case 1, where the original image already has high entropy and pixels evenly scattered, and in case 2-4 where the histogram is squeezed within a small range and is mean shifted in the wide gray scale range available.

Table 2. Entropy, Horizontal and Vertical Correlation values of some Special Test Cases.

| Case | Image | Histogram | Entropy |
| | | | Horizontal Correlation |
| | | | Vertical Correlation |
| 1 | 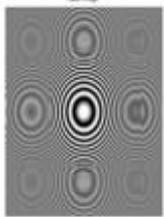 | 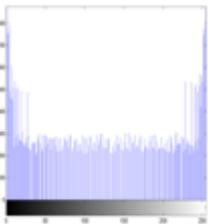 | 7.997100 |
| | | | 0.004073 |
| | | | 0.002616 |

| Case | Image | Histogram | Value |
|---|---|---|---|
| 2 | | | 7.997435 |
| | | | 0.000713 |
| | | | -0.002101 |
| 3 | | | 7.997684 |
| | | | 0.001263 |
| | | | -0.000059 |
| 4 | | | 7.997348 |
| | | | -0.000500 |
| | | | -0.000701 |

Also Table 3 below shows the application of the same on red, green and blue channels of Lena, shown in cases 1-3 respectively and on a binary image case 4. The correlation and entropy values in Table 2&3 prove the effectiveness of the proposed approach.

Table 3. Algorithm Tested for RGB and Binary Images.

| Case | Image | Histogram | Encrypted | Histogram | Entropy<br>Horizontal Correlation<br>Vertical Correlation |
|---|---|---|---|---|---|
| 1 | | | | | 7.999372<br>0.000521<br>-0.002127 |
| 2 | 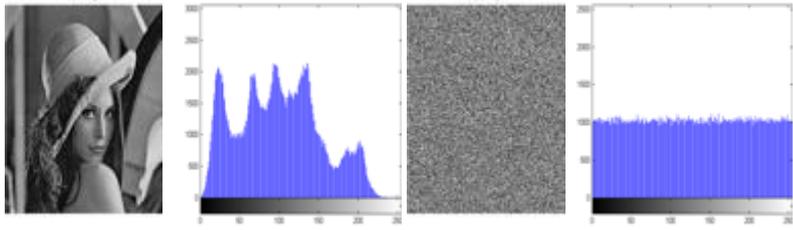 | | | | 7.999316<br><br>-0.000348<br><br>0.002087 |
| 3 | 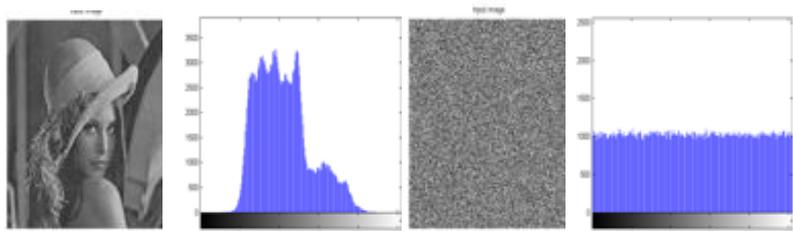 | | | | 7.999418<br><br>0.002986<br><br>-0,000743 |
| 4 | 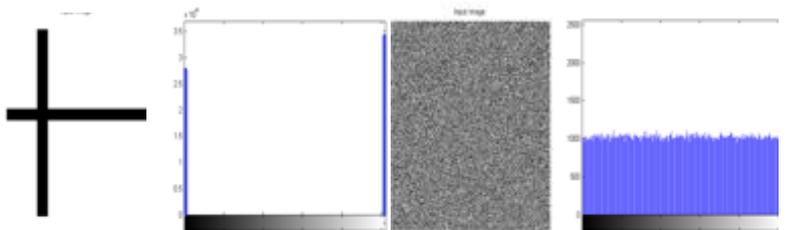 | | | | 7.999365<br><br>-0.002221<br><br>0.001245 |

## 4. Conclusion

In this paper, a new improved approach for image security using a combination of image transformation and encryption techniques is proposed. The approach uses concept of uniform scrambling, row, column and block based image shuffling to reduce correlation. Also double encryption is performed on the shuffled image using a chaotic mapping to enforce more security. For simpler applications user may use the first two steps of encryption process which should be fairly resilient against possible attacks. But in the present age, the proposed approach can be implemented in a multi-core and multiprocessor environment using concepts of pipelining and threading, hence saving on computational time. The experimental results show that the proposed image encryption system has a very large key space; also the cipher image has entropy information close to the ideal value 8 and low correlation coefficient close to the ideal value 0. Thus the analysis proves the security, correctness, effectiveness and robustness of the proposed image encryption algorithm.